\documentclass{aastex}
\usepackage{emulateapj5,apjfonts,psfig}

\def\asca{{\sl ASCA }}

\def\exo{{\sl EXOSAT }}
\def\ein{{\sl Einstein }}

\def\xte{{\sl RXTE }}

\def\chandra{{\sl Chandra }}
\def\ergsec{\hbox{erg s$^{-1}$ }}
\def\ergcm{\hbox{erg cm$^{-2}$ s$^{-1}$ }}

\def\Msun{$M_{\odot}$ }

\def\it{\sl}

\slugcomment{Submitted for publication in The Astrophysical Main Journal}

\shorttitle{HETGS SPECTRA OF CYG X-1}

\shortauthors{SCHULZ ET. AL.}

\begin{document}

\title{The First High Resolution X-ray Spectrum of Cyg X-1: Soft X-Ray Ionization and Absorption}
\author{
N. S. Schulz\altaffilmark{1},
W. Cui\altaffilmark{2},
C. R. Canizares\altaffilmark{1},
H. L. Marshall\altaffilmark{1},
J. C. Lee\altaffilmark{1},
J. M. Miller\altaffilmark{1}
and
W.H.G. Lewin\altaffilmark{1}
 }
\altaffiltext{1}{Center for Space Research, Massachusetts Institute of Technology,
Cambridge, MA 02139.}
\altaffiltext{2}{Department of Physics, Purdue University, West Lafayette, IN 47907.}

\begin{abstract}
We observed the black hole candidate Cyg X-1 for 15 ks with the High-Energy
Transmission Grating Spectrometer aboard the {\em Chandra X-ray 
Observatory}. The source was observed during a period of intense flaring 
activity, so it was about a factor of 2.5 brighter than usual, with a 0.5-10 keV (1-24 ~\AA~) 
luminosity of $1.6\times10^{37}$ \ergsec (at a distance of 2.5 kpc). 
The spectrum of the source shows prominent
absorption edges, some of which have complicated substructure. We use
the most recent results from laboratory measurements and  
calculations to model the observed substructure of the edges. From the
model, we derive a total absorption column of 
6.21$\pm 0.22 \times 10^{21}$ cm$^{-2}$. Furthermore, the results indicate
that there are $\sim 10 - 25\%$ abundance variations relative to solar values 
for neon, oxygen and iron. The X-ray continuuum is described well by a
two-component model that is often adopted for black hole candidates: a 
soft multicolor disk component (with $kT = 203$ eV) and a hard 
power law component (with a photon index of $\sim$2). Comparing the 
fit results to those of the hard and soft states, we conclude that the source 
was in a transitional state. Finally, the spectrum also shows 
the presence of faint emission lines which could be attributed to highly 
ionized species. 
\end{abstract}

\keywords{
stars: individual (Cyg X-1) ---
stars: black holes ---
X-rays: stars ---
X-rays: photo-electric absorption
binaries: close ---
accretion: accretion disks ---
accretion: winds ---
techniques: spectroscopic}

\section{Introduction}

Cygnus X-1 was the first X-ray source to be identified with a binary system
and where radial velocity measurements indicated 
a mass of the compact companion large enough to be a black hole rather than
a neutron star (Webster and Murdin 1972, Bolton 1972). 
The period of the binary was determined optically to be about 5.6 days. 
The system consists of an O9.7 Iab supergiant with a mass in excess of 
$\sim 20$ \Msun in orbit with a black hole of a mass in excess of 
$\sim 7$ \Msun (see Tanaka and Lewin  (1995) for a detailed review).
Cyg X-1 is therefore intrinsically different 
from the majority of known black hole candidates (BHCs) whose stellar components, 
where known, are much less massive. Nevertheless, Cyg X-1 is generally 
regarded as an archetypical BHC; its spectral and temporal properties are 
often used to identify other BHCs, although so far the 
only direct evidence for the black hole nature comes from the dynamical
determination of its mass via radial velocity measurements. The X-ray emission from 
Cyg X-1 is likely due to the release of gravitational
energy of matter that is accreted by the black hole from the companion star.
The mass accretion process is thought to be mediated by a so-called 
"focussed wind"
from the companion star that is close to but not quite filling
its Roche lobe (Gies and Bolton 1986). Tentative evidence for wind accretion
is provided by the observed orbital modulation of X-ray and radio emission
from Cyg X-1 (Kitamoto et al. 2000; Wen et al. 1999; Brocksopp et al. 1999a; Priedhorsky et al. 
1995; and Holt et al. 1979).

In the soft ($<$ 10 keV) X-ray band, the X-ray spectrum of Cyg X-1 can be 
described roughly by a 
two-component model: a multicolor disk component and a power law component. 
The former accounts for the X-ray emission from the hot inner portion of
an optically thick, geometrically thin accretion disk, while the latter
is thought to be due to inverse Comptonization of soft photons (e.g., 
from the accretion disk) by energetic electrons in an optically thin
configuration. The exact shape of the continuum defines two distinct 
spectral states for Cyg X-1: the hard and the soft state. The source is 
usually found in the hard state, where the spectrum is dominated by the
power law component with a photon index of $\sim$ 1.7 (see review
by Tanaka \& Lewin 1995, and references therein). The detection of the
weaker disk component is difficult during this state. The most reliable results came from
the observations of the source with {\em ASCA} (Ebisawa et al. 1996),
which gave a temperature of the disk of only about kT = 100 eV.

Occasionally, for yet unknown reasons, Cyg X-1 experiences a transition
to the soft state, where the disk component grows stronger and the power
law component steepens to a typical photon index of about 2.5 (Tanaka
\& Lewin 1995). The most recent spectral state transition occurred in
1996 (Cui et al. 1996). The source was observed simultaneously by {\em 
ASCA} and {\em RXTE} near the end of the hard-to-soft transition. The
results indicated that the disk component was indeed dominant below 10 
keV and the temperature of the disk was around $kT = 400$ eV (Cui et al. 1998).

The detection of an iron K $\alpha$ line and/or absorption edge has been
reported for Cyg X-1. For instance, a broad emission line seen at 
$\sim$6.2 keV was interpreted to be a red-shifted and broadened
K $\alpha$ line of cold iron (Barr, White and Page 1985; Tanaka 1991;
Done et al. 1992). A narrow emission line was also detected at
6.4 keV in the hard state, which is thought to originate in the 
fluorescent emission from the cold part of the accretion disk (Kitamoto 
et al. 1990; Tanaka 1991; Ebisawa et al. 1992; Ebisawa et al. 1996). A
similar emission line was also seen near the soft state (during the 1996 
state transition) but at $\sim$6.6 keV, which was interpreted as 
emission from highly ionized iron (Cui et al. 1998). 
Iron K $\alpha$ line emission has become one of the most useful tools
for probing the strong gravitational field near a black hole. The 
detection of a gravitationally distorted line would provide strong 
evidence for the existence of a black hole. Such a line has been 
reported for the Seyfert 1 galaxy MG-6-30-15 (Tanaka et al. 1995), but 
has never been detected in Cyg X-1.

In this paper we present the first high resolution X-ray spectra of 
Cyg X-1 obtained with the High Energy Transmission Gratings Spectrometer
(HETGS, Canizares et al. 2001, in preparation) onboard the \chandra X-ray Observatory (\chandra hereafter) 
during its early phase of the mission. In this paper we take a fresh look
at the soft part of the spectrum below 10 keV, particularly 
below 1 keV. We report the detection of prominent absorption edges
in the spectrum of Cyg X-1, as well as weak emission lines due to 
highly ionized species.
  
\section{Chandra Observations and Data Reduction}

Cyg X-1 was observed with the HETGS on 1999 October 15 (start time 05:47:21 UT) 
continuously for 15 ks.
For a general description of the HETGS we refer to available Chandra 
X-ray Center (CXC)
documents \footnote{http://asc.harvard.edu/udocs/docs/docs.html}. 
In this section we outline the overall properties
of this observation, which was performed using the Advanced CCD Imaging Spectrometer (ACIS, Garmire et al. 2001,
in preparation) in standard timed event mode,
but with alternating frame times to accomodate the high source flux.

The HETG disperses about half of the incident photons into 1st and higher order spectra, which are recorded
with an array of 6 CCDs at the focal plane of the telescope. All other
photons generate an image in the zeroth order, which is located at the focus of the telescope.
Every set of seven CCD frames of 3.3 sec was followed by a short frame of
0.5 sec. This resulted in an effective exposure of 11.5 ks in long frame mode and
0.3 ks in short frame mode. Separate event lists were delivered by the CXC standard
processing. Throughout this paper, we distinguish between a long frame observation 
and a short frame observation, although the reader should always keep
in mind that both observations were performed during the same time period. In order to avoid frame dropouts due to
telemetry saturation, we blocked the zeroth order events electronically from being transmitted. Therefore standard
source detection is not possible and we ignored any products from standard processing
higher than level 1. 

The determination of the zeroth order position is crucial for the calibration
of the wavelength scales because this position defines its zero point. We calculated this position
by fitting the dispersed images of the Medium Energy Gratings (MEG) and High Energy Gratings (HEG) 
and then determined the intersections of the 
fit with the zero order readout trace of the CCD. The intersections for the MEG and HEG were determined to be within
0.2 detector pixels. From this, we deduce an accuracy of the zero order position of 0.002 \AA ~in MEG and 0.001 \AA ~in HEG
1st order. This uncertainty is negligible when compared to the overall wavelength calibration, 
which has been verfied on orbit to be better than $\sim$ 0.05$\%$ (Canizares et al. 2001, in preparation).
The worst case uncertainty in the wavelengths then is 0.011 \AA~ in the MEG 1st order near
the O K edge, 0.008 \AA ~ in the HEG 1st order near the Ne  K edge and 0.003 \AA ~
in the MEG 3rd order near the Mg K edge.
We processed the event lists into grating (level 1.5) event lists using available CXC soft
ware. From there on we used the Chandra Interactive Analysis of Observations package (CIAO 2.1\footnote{http://cxc.harvard.edu})
and custom software to produce our final grating spectra. 
We computed aspect corrected exposure maps for each spectrum which allow us to calculate 
the effective area of the spectrometer to obtain absolute fluxes. Current systematic uncertainties
of the effective area calibration are $\sim$ 10 to 20$\%$; in some cases we were able to 
correct for local flux enhancements by comparing opposite sides of the dispersion. 

High X-ray fluxes cause photon pileup in X-ray CCD devices. In the case of Cyg X-1 the flux is
so high that photons pile up even in the grating spectra during long-frame exposures. Pileup occurs when
two or more photons of the same energy are registered almost simultanously and thus identified
by the CCD device as one photon of the summed energies of the registered photons. In the case of the
grating spectra this means that photons from 1st order spectra are registered in higher CCD channels
which are coincidently identified for photons form higher order spectra. The 1st order 
spectra below $\sim$ 14~\AA~ (MEG) and  $\sim$ 10~\AA~ (HEG) are depleted of photons, which then
populate the second and third order spectra at wavelengths below half and one third of the wavelengths,
respectively. The MEG 3rd order spectrum below $\sim$ 4.6~\AA~
is therefore 'contaminated' by photons piled up from MEG 1st order.

\section{Light Curve}

\centerline{\epsfxsize=8.5cm\epsfbox{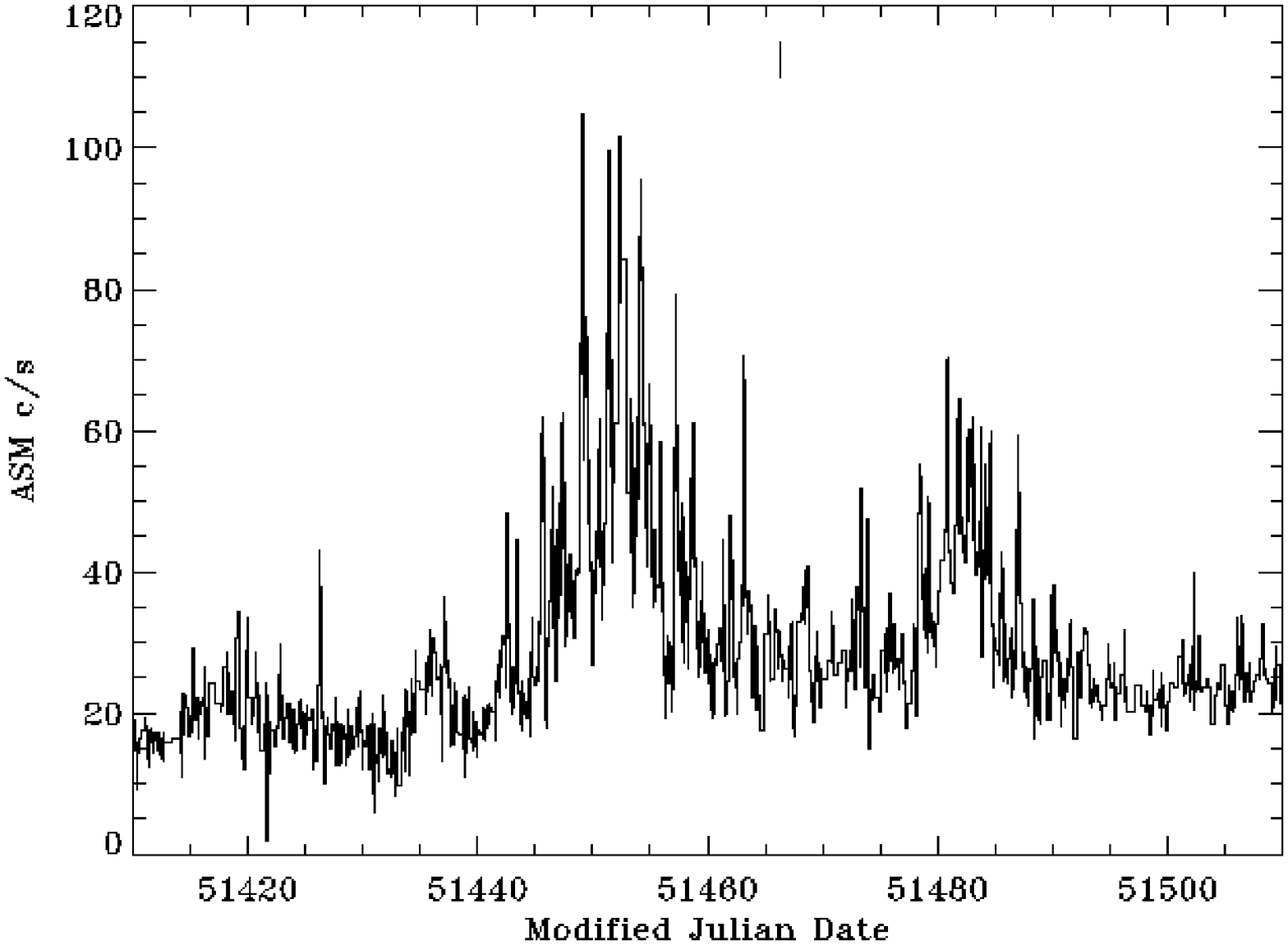}}
\figcaption{ASM light curve of Cyg X-1 of the soft flares in the vicinity of the
HETGS observation. The vertical line at the top indicates the time of
the {\em Chandra} observation.
\label{nek}}
\vspace{2mm}

Cyg X-1 was observed during a period when it produced frequent X-ray
flares. Flaring activity is common for the source, as shown by the
long-term monitoring of the source by the All-Sky Monitor (ASM) aboard 
{\em RXTE} (see Fig. 1). However, X-ray flares are usually brief, and are therefore
difficult to catch in short observations. Consequently, they are poorly understood due to the
lack of data. The {\em Chandra} observation now provides us with a rare 
opportunity to examine the properties of Cyg X-1 during such an event.
Figure 1 shows a portion of the ASM light curve for a time span of about
100 days during which the short (15 ks) Chandra observation was performed. The time of the
{\em Chandra} observation is indicated in the figure. As can be seen, 
Cyg X-1 was engaged in heavy flaring activity and the 
observation lies in between two giant flares. We also note
that there are gaps in the ASM coverage. 
Some brief mini-flares could therefore have been missed during the ASM coverage.
The {\em Chandra}
light curve is difficult to analyse because the source appeared at a rate of over 200 cts per sec in
the HETGS 1st orders and light curves are systematically affected when we include these data. 
Excluding all grating data affected by pileup (with a confidence level of better than 99.9$\%$), 
we obtained a light curve for the bandpass above 
6 ~\AA~, which showed variability of roughly 20$\%$ over the  
observation. 

\section{Spectral Analysis}

\centerline{\epsfxsize=8.5cm\epsfbox{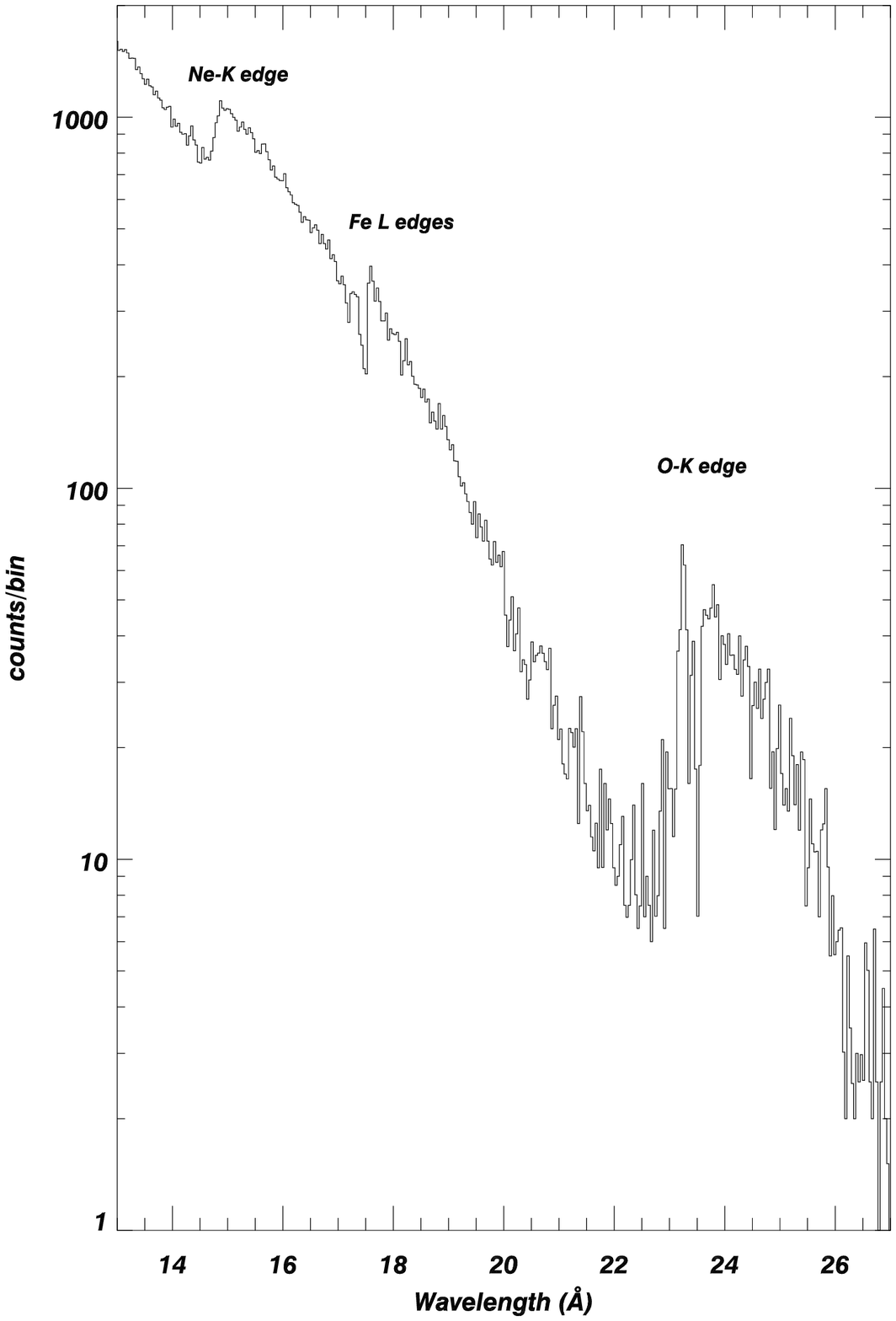}}
\figcaption{The MEG spectrum in the 13 - 27 ~\AA~ band, where it is unaffected by pileup. The binsize is 0.04 ~\AA~.
The data are not corrected for instrumental effects.
\label{data}}
\vspace{2mm}

Major portions of the 1st order grating spectra
were severely piled up and only of limited quality and not all wavelengths can be used to 
to study the broad band continuum shape in the long frame observation.
In the following analysis, we 
look in detail at the observed neutral edges from which we are able to derive
reliable column densities of the absorbing material. We use these column densities to fit the spectral continuum
in the subsequent section.
Using the optical depth of each detected edge from the spectrum,
we derive the column density for that particular 
element. Using the combined results for various elements, we then determine
an overall column density, as well as abundance variations of the elements in the absorbing gas.
Finally we identify residual absorption and emission features. 

\subsection{Photoelectric Absorption}

In low resolution detectors the detailed shape of photo-electric absorption edges
were never resolved and therefore were largely ignored. In a high-resolution spectrum
of the Crab Nebula 
using the Focal Plane Crystal Spectrometer onboard \ein Schattenburg and Canizares (1986)
presented the first detailed X-ray spectrum near the O K edge and 
found marginal evidence for a resonant $1s-2p$ absorption line from atomic oxygen.
Paerels et al. (2001) showed with 
\chandra Low Energy Transmission Grating data near the neutral oxygen edge of another X-ray binary, 
4U 0614+091, there is considerable structure due to absorption lines from
atomic oxygen as well as other molecular compounds. For the case of Cyg X-1,
we observe similar complex substructures which we quantify
by comparing them with expected structure obtained from laboratory measurements
as well as from calculations.  

The MEG count spectrum (Fig. 2)
shows the photo-electric absorption edges of O K, Fe L, and Ne K of Cyg X-1.
A detailed study of the structure
of photoelectric absorption in observations of a sample including other bright X-ray binaries will be presented 
by Schulz et al. (2001, in prep).

Neutral absorption column densities are directly proportional to the optical depth of the neutral
edge. The latter is defined by 

\begin{equation}
\tau = ln (f_{high} / f_{low}),
\end{equation}

\noindent
where f$_{high}$ is the X-ray flux on top of the edge at the high wavelength side
and f$_{low}$ is the X-ray flux at the bottom of the edge at the low wavelength side.
We emphasize that this optical depth corresponds to the total amount of photo-electric absorption,
which includes the interstellar medium in the line of sight as well as a possible intrinsic contribution
by the source itself.  
We observe edges at O, Fe, Ne, and Mg. The data allow to directly measure the
optical depths in the cases of O, Fe, and Ne. The Mg edge is visible but the bins for f$_{high}$
and f$_{low}$ have large uncertainties. The Si edge does not appear significantly in 
the data.    
Table 1 shows the expected and measured values of the edges and the measured optical depths from these edges.
We quadratically added a systematic uncertainty of one spectral resolution bin to the wavelength uncertainty, which
also dominates the uncertainty of the edge value. 

\begin{table*}
\footnotesize
\begin{center}
{\sc TABLE 1\\
POSITIONS AND OPTICAL DEPTHS FROM PHOTO-ELECTRIC EDGES}
\vskip 4pt
\begin{tabular}{lccccc}
\tableline
\tableline
     edge & $\lambda_o$ & $\lambda_m$ &$\tau_m$& NH$_z$& a$_z^5$ \\
         & \AA & \AA & &  cm$^{-2}$& \\
\tableline
     & & & & & \\
     O K         & 23.34$^1$ & 23.15$\pm0.02$ & 2.226$\pm0.133$ & 3.92$\pm 0.23 \times 10^{18}$ & 0.93 \\
     Fe L 3      & 17.51$^2$ & 17.56$\pm0.02$ & 1.098$\pm0.086$ & 1.55$\pm 0.12 \times 10^{17}$ & 0.75 \\
     Fe L 2      & 17.19$^2$ & 17.23$\pm0.02$ & 0.307$\pm0.027$ & 1.46$\pm 0.12 \times 10^{17}$ &  -   \\
     Fe L 1      & 14.66$^1$ & 14.66$\pm0.02$ & 0.079$\pm0.017$ &  - & - \\
    pure Fe L    & 17.44$^3$ & 17.41$\pm0.02$ & 0.058$\pm0.012$ & 8.24$\pm 1.72 \times 10^{15}$ & - \\
    Fe$_x$O$_y  $& 17.64$^4$ & 17.66$\pm0.02$ & 0.025$\pm0.009$ & 4.49$\pm 1.61 \times 10^{15}$ & - \\
     Ne K        & 14.30$^1$ & 14.32$\pm0.02$ & 0.343$\pm0.011$ & 9.43$\pm 0.32 \times 10^{17}$ & 1.11 \\
     Mg K        &  9.50$^1$ &  9.49$\pm0.02$ & 0.088$\pm0.032$ & 3.67$\pm 1.34 \times 10^{17}$ & 1.00 \\
     Si K        &  6.74$^1$ &  6.74$\pm0.01$ & 0.036$\pm0.028$ & 2.31$\pm 1.81 \times 10^{17}$ & 1.00 \\
     & & & & \\
\tableline
\vspace*{0.02in}
\end{tabular}
\parbox{5.5in}{
\small\baselineskip 9pt
\footnotesize
\indent
$\rm ^1${from Bearden $\&$ Burr 1967}, $\rm ^2${from Kortright $\&$ Kim 2000}, $\rm ^3${from Brown 2000}, $\rm ^4${from Crocombet
te et al. 1995}
$\rm ^5$ abundance adjustment factor
}
\end{center}
\setcounter{table}{1}
\normalsize
\centerline{}
\end{table*}

The edges exhibit considerable substructure, which we describe in detail below.
We first determined the optical depths $\tau$ of the O-K, Fe-L, and Ne-K edges using equation 1.
The optical depth ratios between these edges showed some deviations to the 
ones expected for abundances derived using the polynomial fit parameters from Morrison $\&$ McCommon (1983) as
implemented in XSPEC using the cross sections from Verner et al. (1993).  
We also applied a continuum model fit derived from the short frame data using
a pre-determined value for $N_H$ derived mainly from a first assessment of the Ne-K optical depth, which shows the
least structure. Assuming a solar abundance relative to hydrogen as stated by Morrison and McCammon (1983) we find a
best value of $N_H$ = 6.21$\pm 0.22 \times 10^{21}$ cm$^{-2}$.  
For elements other than Ne, O, and Fe we use the solar abundances relative to hydrogen. In the case of 
Ne, O, and Fe we use the long frame data to tweak the optical depths by adjusting abundances. The abundance adjustment factor
$a_z$ is shown in Table 1. For O and Fe we
find a lower abundance of $\sim$ 7$\%$ and  $\sim$ 25$\%$, respectively. For Ne we find a higher abundance of 11$\%$.    
From the final adjusted models we then determined the optical depths $\tau_m$ shown in table 1. 
We have also calculated the column density for each ion species $Z$, defined as 

\begin{equation}
N_z = {\tau_{m} \over \sigma_z},
\end{equation}

\noindent
where $\sigma_z$ are the atomic photoelectric cross sections. We use mostly the cross sections from Verner et al. 1993, except
for Fe, where we have new cross sections from Kortright $\&$ Kim (2000).
The final spectral model was generated by computing the absorption edges using $\tau_{m}$ and also by adding single gaussians
for local absorption dips.

\centerline{\epsfxsize=8.5cm\epsfbox{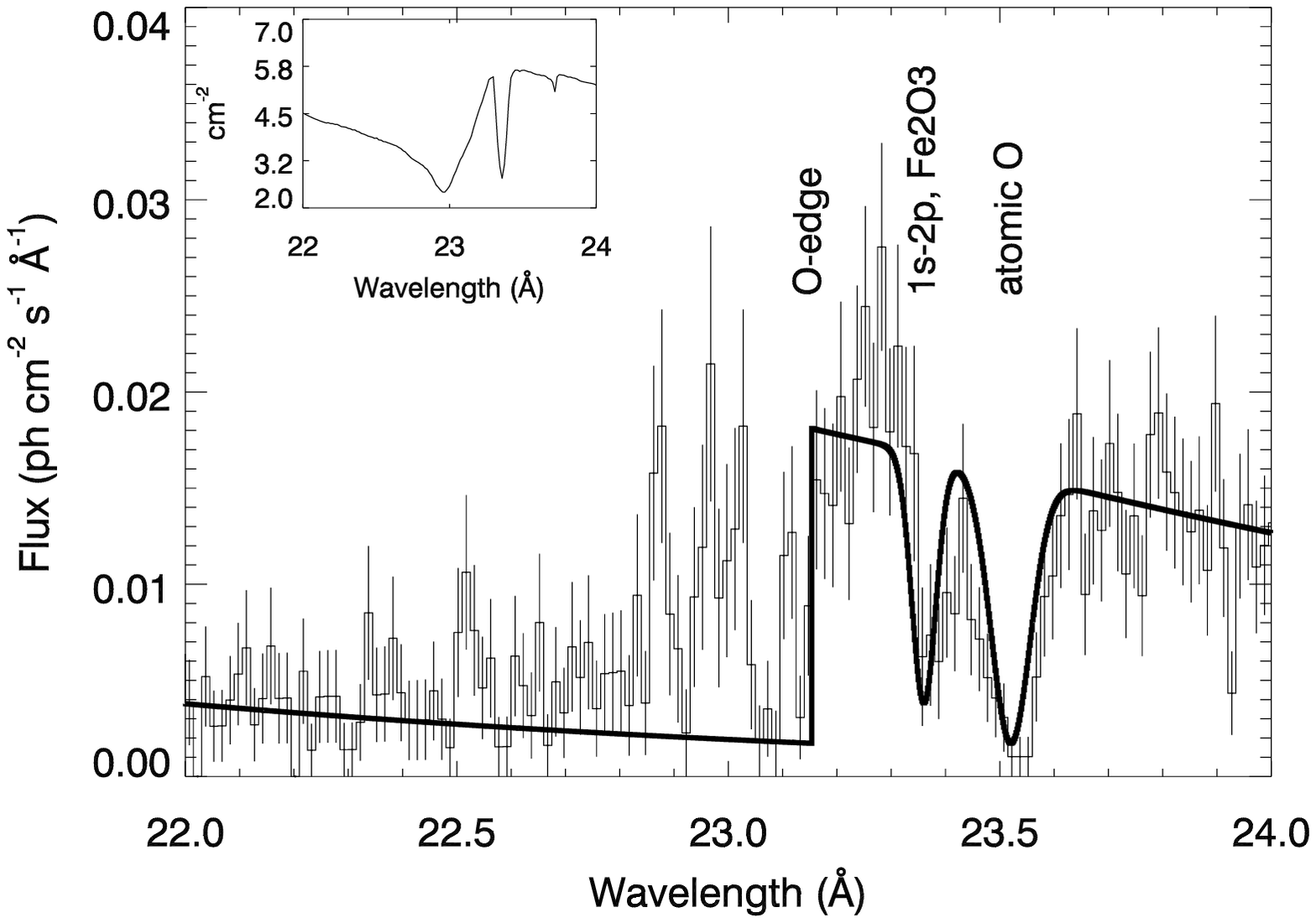}}
\figcaption{A MEG view of the O K edge region. The curve represents the model. The inset shows
the effective ares of the MEG around the O K edge.
\label{ok}}
\vspace{2mm}
 
\subsubsection{The O K edge}

Structure in the O K edge is difficult to assess with the MEG because the effective area
of the instrument is relatively low and affected by instrumental O features. 
The source is easily detected longward of the edge (see Fig. 2);
statistical uncertainties in the data are of the order of $<30\%$ (Figure 3). Shortward of the edge, the spectrum
is very faint. The most probable position of the O K edge appears at 23.15 ~\AA, which is lower than the expected
value of 23.31 ~\AA~ (Bearden and Burr 1967), but consistent with a most recent observation
of Cyg X-1 (Marshall et al. 2001). The value of the O-edge in this observation is difficult to assess, because  
we observe excess emission around 23 ~\AA. We consider two likely 
interpretations. First, the excess emission would result from a systematic error in the effective area model (see inset in Fig. 3), 
which is dominated by the O K edges of polyimide in the ACIS filter and the MEG support material. Based on pre-launch
calibration, however, systematic errors in the HETGS effective area should be no more than 50$\%$, while an error 
four times or more would be required to eliminate the feature at 23~\AA~. 
Second, we cannot rule out the possibility that some of the excess emission consists of emission lines from Cyg X-1 
(see section 4.3). Given the results by Marshall et al. 2001, the latter interpretation is more likely.   
The feature at 23.36 ~\AA~ coincides with a resonance feature in polyimide,
but there is a significant residual, which may be identified with the 1s-2p transition
from oxygen in Fe$_2$O$_3$ (Wu et al. 1997). Most prominently the large absorption dip with a FWHM of 
0.08 A at 23.51 ~\AA~ due to the 1s-2p transition in atomic oxygen. This feature was marginally detected by Schattenburg and Canizares 
(1983) in a Crab Nebula observation, but appeared prominently in the LETG observation of X 0614+091 (Paerels et al. 2001).

\centerline{\epsfxsize=8.5cm\epsfbox{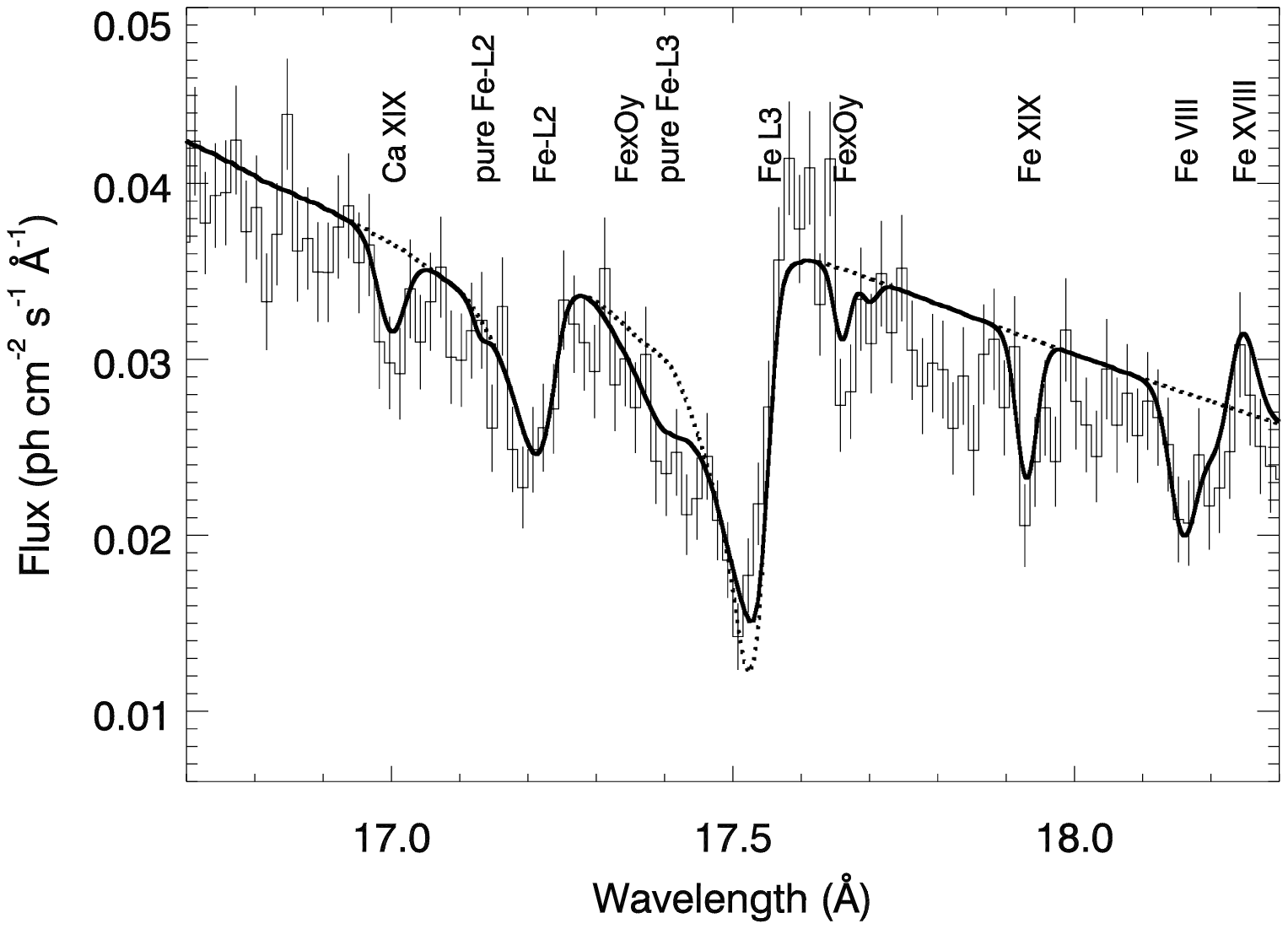}}
\figcaption{A portion of the MEG spectrum showing the photo electric Fe L III and II edges. The solid curve is the final mo
del,
the dotted line shows the shape of the Fe 3 and 2 edges from Kortright and Kim (2000).
\label{fek}}
\vspace{2mm}

\subsubsection{The Fe L edges}

Iron is the most abundant element that has its L absorption in the HETGS
bandpass. The Fe-L1 edge is due to transitions from the 2s level, Fe-L2 from the 2p$_{1/2}$, and Fe-L3 from the 
2p$_{3/2}$ level. The Fe-L3 edge is usually strongest, the Fe-L1 edge is weakest. Figure 4 shows the Fe-L
edge region in the MEG spectrum. We clearly observe that these edges have narrow absoption features, which cannot be
described by a single step function. The Fe-L1 edge region is shown in Figure 5. The shapes of the Fe-L edges were
measured by various groups using metallic iron of various purity as well as iron oxides (Kortright $\&$ Kim 2000, Brown 2000,
Schwickert et al. 1998, Crocombette et al. 1995). 
These studies indicate that the position of the Fe L edges depend on whether the sample is oxidized of metallic.
We measured the
position of the Fe L3 edge (defined to be the top of the edge) to be 17.56 ~\AA~. 
This position is consistent with recent measurements of 
metallic Fe film by Kortright $\&$ Kim (2000), but is between results from Fe foils of very high purity
(Brown 2000) and Fe oxides (Crocombette et al. 1995). We used new experimental cross sections from
Kortright$\&$ Kim (2000) in the absorption model shown in Figure 4.
This model represents the overall observed shape of the edges quite well, with some  
caveats. First of all, the column density of Fe is $\sim 25\%$ lower than expected
from the Morrison $\&$ McCommon representation. We observe some slight deviations from the shape predicted 
by Kortright$\&$ Kim (2000), which may indicate that Fe exists in various forms,
or that the edge has superimposed intrinsic emission and absorption lines.  
We observe additional, but very weak features around 17.41 ~\AA~ and 17.66 ~\AA~, which match the
positions that Brown (2000) determined from an extremely pure Fe sample as well as those from 
Fe oxides (Crocombette et al. 1995), respectively.
We added absorption from these forms of Fe into the model assuming the same shape of the cross section as from the Kortright data.
With this approach we would recover about $7\%$ of the missing Fe. Wilms, Allen and McCray (2000) list 
an amount comparable to the remaining 18$\%$ as the difference between the solar abundance of Fe and 
the one in the interstellar medium.       
There are also faint features from ionized species (see section 4.3), which appear
in emission as well as absorption. From our analysis in section 4.3 it seems likely that these distortions 
are due to emission and absorption from highly ionized iron, so we simply added Gaussian line components to the model. The normalizations
were determined by fitting single Gaussians to the data, while the widths were fixed to the MEG spectral resolution.   

\subsubsection{The Ne K-, Mg K-, and Si K edges}

\centerline{\epsfxsize=8.5cm\epsfbox{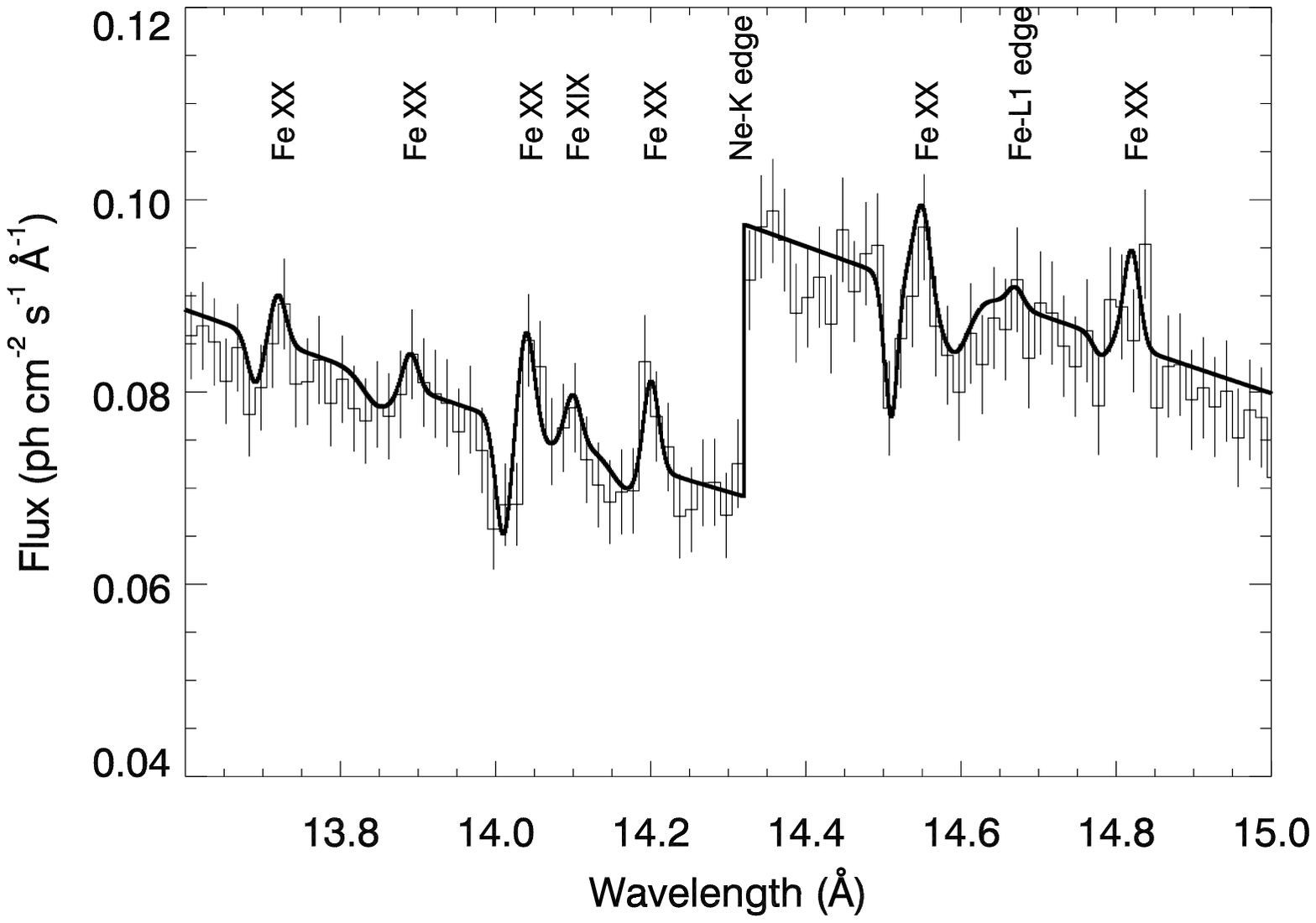}}
\figcaption{A portion of the MEG spectrum which shows the photoelectric edge of Ne K and numerous weak ionized iron lines.
\label{nek}}
\vspace{2mm}

The Ne K edge (see Figure 5) can be very well modeled using the Verner et al. (1993) cross section,
since it appears only in gaseous form.
The wavelength has been measured in the laboratory to 14.302$\pm$0.003 ~\AA~ in agreement 
with the calculated value of 14.30 ~\AA` (Bearden and
Burr 1967). We measure a value of 14.32$\pm$0.02 ~\AA~, which is consistent with the expected value above. 
It is also consistent with the value observed in X0614+091 (Paerels et al. 2001), which however has a very large
uncertainty. 
There is considerable substructure in the vicinity, which we identify as emission lines (see section 4.3) similiar
to the features in the vicinity of the O and Fe edges.

The Mg K and Si K edges (see Figure 6) appear at wavelengths where the MEG and HEG spectra are
significantly piled up and we have to rely on the spectrum from the MEG 3rd order, which has an order of magnitude lower
efficiency. Both edges are modeled using the Verner et al. (1993) cross sections and the abundance factor $a_z$ in Table 1
is set to 1.

\centerline{\epsfxsize=8.5cm\epsfbox{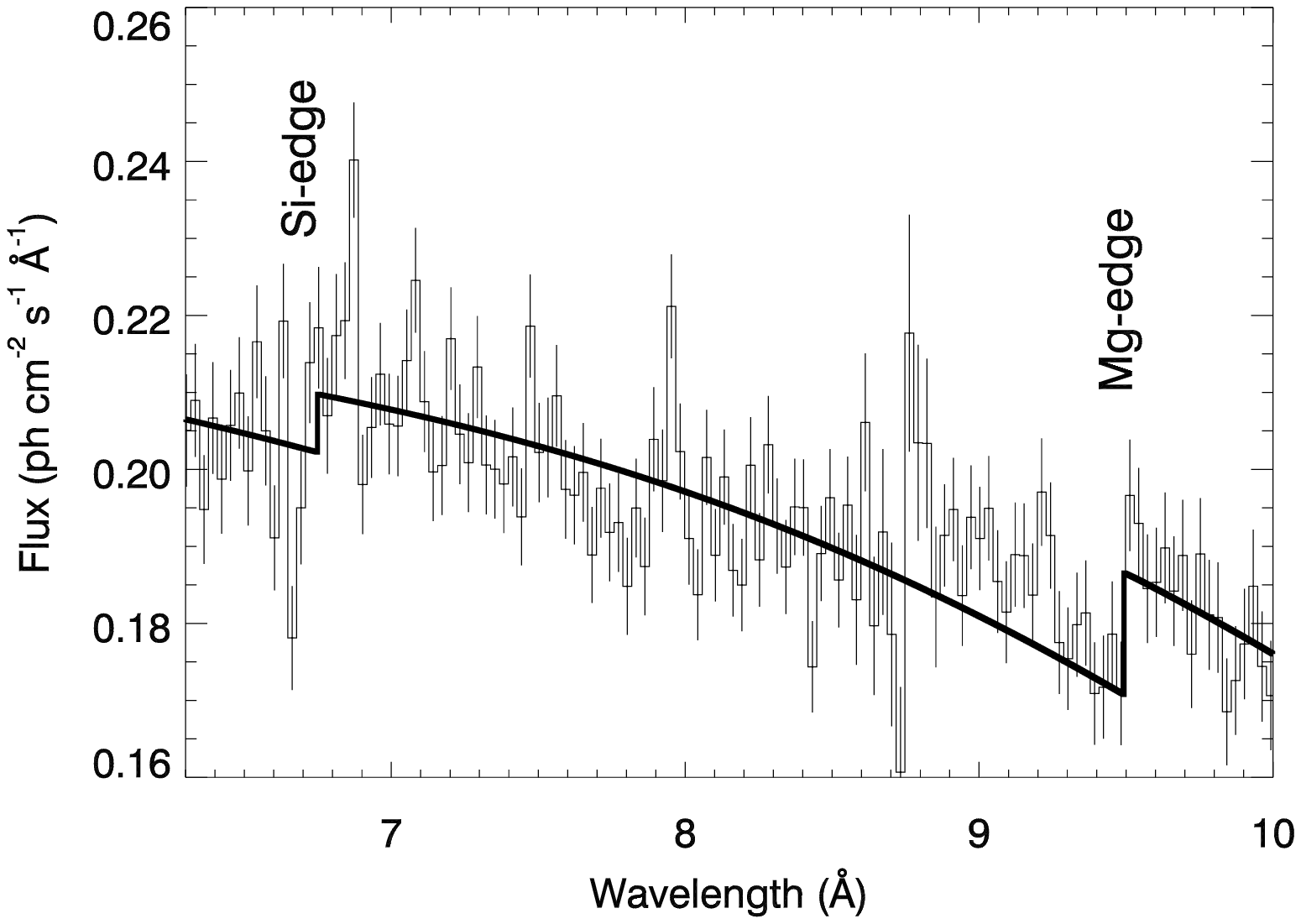}}
\figcaption{A portion of the MEG 3rd order spectrum which shows the photoelectric edges of Mg K and Si K.
\label{fek}}
\vspace{2mm}

\subsection{The X-ray Continuum}

For the determination of the continuum spectral shape we use the short as well as
the full frame observations. The short frame data provide us with a pileup free
continuum at the expense of exposure. We therefore determined the 
continuum in two steps. First we fit the spectra from the short frame observation using the XSPEC package
in order to determine the overall spectral shape and also define the spectral
model to fit the continuum. Second, we test the fit on a 
spectrum obtained by combining pileup-free portions of the long frame observations, where
we combine portions from the MEG 1st (14-26~\AA~), HEG 1st (10-14~\AA~), and MEG 3rd (6-10~\AA~) order. Both, short frame
and long frame  spectra,
have their limitations: the short frame spectra have large statistical uncertainties,
the long frame data are statistically very significant but only accurate for wavelengths $>$ 6 ~\AA~.

\begin{figure*}
\centerline{\epsfxsize=17.5cm\epsfbox{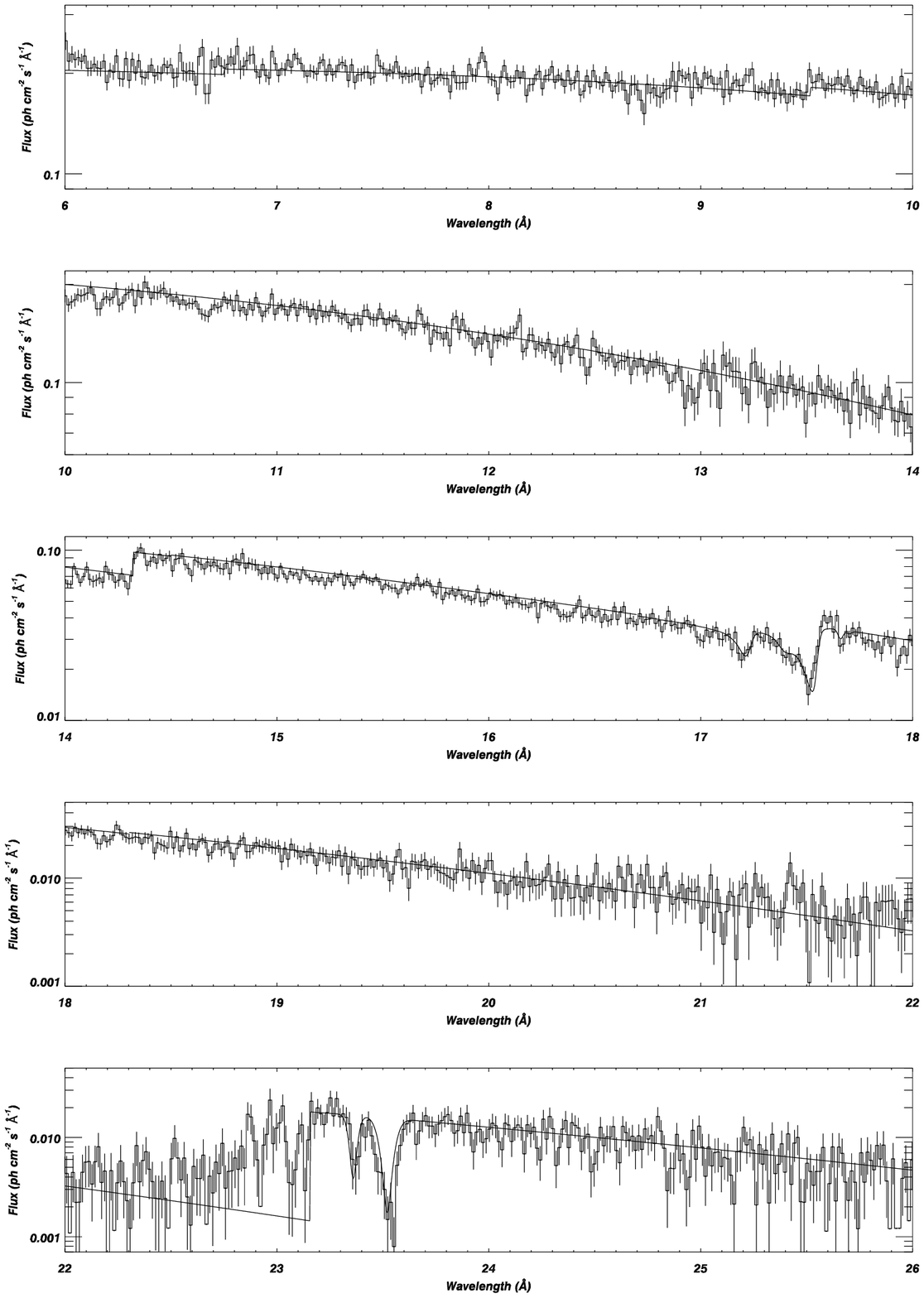}}
\figcaption{The broadband HETGS spectrum between 6 and 26 ~\AA~. The continuum model was fitted with a disk blackbody
plus power law model. The lower three panels are from the MEG 1st order spectrum, the second one from top
shows a portion of the HEG 1st order spectrum, the top panel shows the MEG 3rd order spectrum.
\label{continuum}}
\end{figure*}

A simple power law does not give a good fit even to the short frame data. A 
second power law component improved the fit to the short-frame data 
considerably, but leaves significant residuals when applied to the long-frame
data. Using the multicolor disk model (as used by Cui et al. 1997) as a second 
component we obtain an excellent fit to the data ($\chi^2 < 1$; 2206 d.o.f). The result, shown in Figure 7, 
also includes the photoelectric absorption models. 
The power law has an index of 1.95$\pm 0.07$, the disk component has a temperature of $kT$ = 203$\pm 4$ eV.   
The total absorbed flux is 1.03$\times 10^{-8}$ \ergcm between 0.5 and 10 keV ($\sim$
1 - 25 ~\AA~), with the disk component contributing about 11$\%$.
Assuming a distance of 2.5 kpc to the source, we determine a luminosity
of the disk blackbody component of 5.34$\times 10^{36}$ \ergsec and
a total luminosity of 1.61$\times 10^{37}$ \ergsec. Compared with
the properties of the source in the hard and soft states (e.g., Ebisawa et al. 1996, Cui et al. 1998),
our results clearly show that the source was
in a transitional state, as expected from the ASM light curve (see Fig.~1).

\subsection{Narrow Emission and Absorption Features}

\vbox{
\footnotesize
\begin{center}
{\sc TABLE 2\\
WEAK AND NARROW LINE FEATURES FOR WAVELENGTHS $>$ 6 ~\AA}
\vskip 4pt
\begin{tabular}{lcccccc}
\tableline
\tableline
    $\lambda_{em}^1$& Ion     &  log T$^2$    & S/N & $\lambda_{abs}^1$ & ion & S/N \\
      \AA         &         & (k) &          &         \AA     &     & \\
\tableline
     & & & & & & \\
          &            &        &        &    6.65   &  Si XIII   &  4.1 \\
     6.88 &  Si X      &  6.3   &   3.4  &           &            &      \\
     7.09 &  Mg XII    &  7.0   &   3.3  &           &            &      \\
     7.31 &  Mg XI     &  6.8   &   2.3  &           &            &      \\
          &  Ni XXV    &  7.2   &        &           &            &      \\
     7.47 &  Mg XI     &  6.8   &   2.1  &           &            &      \\
          &  Fe XXIII  &  7.2   &        &           &            &      \\
     7.68 &  Fe XXIII  &  7.2   &   -    &           &            &      \\
     7.98 &  Fe XXIV   &  7.3   &   3.2  &           &            &      \\
          &            &        &        &    8.05   &  Fe XXII   &  2.1 \\
          &            &        &        &    8.19   &  Fe XXII   &  3.0 \\
          &            &        &        &    8.42   &  Mg XII    &  2.3 \\
          &            &        &        &    8.72   &  Fe XXIII  &  2.8 \\
     8.76 &  Fe XXIII  &  7.2   &   2.8  &           &            &      \\
     9.22 &  Mg XI     &  6.8   &   4.3  &           &            &      \\
          &            &        &        &   10.15   &  Fe XX     &  3.1 \\
          &            &        &        &   10.65   &  Fe XXIII  &  5.3 \\
    11.67 &  Fe XVIII  &  6.8   &   2.3  &           &            &      \\
    11.84 &  Fe XXII   &  7.1   &   2.3  &           &            &      \\
    11.88 &  Fe XVIII  &  6.9   &   3.3  &           &            &      \\
          &            &        &        &   12.01   &  Fe XXI    &  2.4 \\
    12.14 &  Ne X      &  6.8   &   2.2  &           &            &      \\
          &            &        &        &   12.48   &  Fe XXI    &  4.2 \\
    12.50 &  Fe XXII   &  7.1   &   2.9  &           &            &      \\
          &            &        &        &   12.94   &  Fe XX     &  4.5 \\
          &            &        &        &   12.99   &  Fe XX     &  4.4 \\
    13.04 &  Fe XXII   &  7.1   &   2.5  &           &            &      \\
    13.14 &  Fe XX     &  7.0   &   2.2  &           &            &      \\
          &            &        &        &   13.50   &  Fe XX     &  2.0 \\
    13.56 &  Ne IX     &  6.6   &   2.1  &           &            &      \\
          &  Fe XX     &  7.0   &        &           &            &      \\
          &            &        &        &   13.69   &  Fe XX     &  2.3 \\
          &            &        &        &   14.02   &  Fe XIX    &  3.1 \\
    14.04 &  Fe XX     &  7.0   &   3.0  &           &            &      \\
    14.20 &  Fe XX     &  7.0   &   3.3  &           &            &      \\
          &            &        &        &   14.49   &  Fe XIX    &  4.7 \\
    14.55 &  Fe XX     &  7.0   &   4.1  &           &            &      \\
    14.83 &  Fe XX     &  7.0   &   2.1  &           &            &      \\
    16.42 &  Fe XX     &  7.0   &   3.0  &           &            &      \\
          &            &        &        &   17.00   &  Ca XIX    &  3.6 \\
          &            &        &        &   17.41   &  Fe XVIII  &  3.0 \\
    17.54 &  Fe XIX    &  6.9   &   3.8  &           &            &      \\
          &            &        &        &   17.93   &  Fe XIX    &  4.3 \\
          &            &        &        &   18.18   &  Fe XVIII  &  4.4 \\
    18.25 &  Fe XVIII  &  6.8   &   2.7  &           &            &      \\
          &            &        &        &   18.81   &     ?      &  2.8 \\
    18.83 &  O VIII$^3$&  6.5   &   2.0  &           &            &      \\
    21.43 &  O VII$^3$ &  6.3   &   3.4  &           &            &      \\
    22.52 &  O V       &  6.2   &   2.7  &           &            &      \\
    22.86 &  O IV      &  5.7   &   4.1  &           &            &      \\
    23.05 &  S XIV     &  6.5   &   3.3  &           &            &      \\
    23.10 &  S XIII    &  6.4   &   3.5  &           &            &      \\
    & & & & & & \\
\tableline
\end{tabular}
\end{center}
\parbox{3.2in}{
\small\baselineskip 9pT
\footnotesize
\indent
$\rm ^1${from SPEX 1999 line list} \\
$\rm ^2${temperature of maximum emissivity}\\
$\rm ^3${possibly blue shifted by 2350 km s$^{-1}$}
}
\setcounter{table}{1}
\normalsize
\centerline{}
}

There are a variety of emission and absorption features in the vicinity of the edges (Figures 3-5).
A few are also visible in figure 7.
Nearly all of these features are weak, detected with a signal to noise (S/N) ratio of
less than 5. The S/N ratio is Table 2 is defined as

\begin{equation}
S/N = {{\sum f^l_i}\over{\sqrt{\sum(\sigma^c_i)^2}}},
\end{equation}

\noindent
where $f^l_i$ is the flux of a line in the spectral bin $i$ and $\sigma^c_i$ the significance of the
continuum in that bin and the sum is over the line width. The flux in each bin is the value
of the bin subtracted by the value of the model in that bin multiplied by the width of the bin.
The uncertainty in each bin of the 
underlying continuum was determined from the square root of the model value folded 
through exposure and effective area.
We exclude all bandpass regions where
pileup effects are significant. The detected emission as well as absorption features
and the positions ($\lambda_{em}$ for emission,
$\lambda_{abs}$ for absorption),
tentative identifications, and detection significances are listed in table 2.
The identifications were made using the
line list from Mewe (1994). Features with a S/N ratio near 2 are not very
significant, but in the context of the ones with higher ratios have valid identifications.
We identify ion species mostly from
intermediate to highly ionized Fe with ionization stages between Fe XVIII and Fe XXIV.
In column 3, we also list the temperature corresponding to the peak emissivity of the transition.
These indicate temperature values between 1.5 and 20 Million K.
Many of the features are seen in the Figures 3 to 5, but
we also show the 11.5 - 13.6 ~\AA~ bandpass from the HEG spectrum in Figure 8. All the features are
narrow and do not appear resolved by the MEG resolution.
The observed flux range is 2.1 - 15 $\times10^{-5}$ ph cm$^{-2}$ s$^{-1}$.

There are several difficulties associated with the identifications of these weak lines. In Figures
3-8 we see that there are also absorption features and many of them seem to be associated
with an adjacent emission line which gives the impression of a P Cygni type line profile, such as has
been observed in Cir X-1 (Brandt and Schulz 2000). We list
those in the right part of table 2. Note, that Marshall et al (2001) also report strong absorption
lines in another HETGS observation of Cyg X-1.
In general, identifications in the wavelength band below  16 ~\AA~
are not entirely unique, because the line density is so high, increasing the likelihood of misidentification.
At longer wavelengths ($> 16$ ~\AA), the Fe transition forest thins so
identifications are more straightforward.

Figure 9 shows a weak detection at 21.43 ~\AA~.
This line coincides with a line from Ca XVI at 21.44 ~\AA~, which given the temperature
distribution in column 3 of table 2, is an unlikely identification in the absense of
other nearby Ca lines. We tentatively attribute this feature
to the O VII He-like resonance line, which would mean
a shift corresponding to about -2350 km s$^{-1}$. Similarly we detect a fainter feature at 18.83 ~\AA~, with some
absorption at 18.81 ~\AA~, which are both not identified.
It could be identified with the O VIII Lyman $\alpha$ transition, if it is also blue-shifted by 2350 km s$^{-1}$.

\section{Summary and Discussion}

The first high resolution X-ray spectrum of Cyg X-1 obtained with \chandra shows various features, which
will be discussed in more detail below. In the following we briefly summarize the main results from the \chandra observations.
From resolved neutral edges of various elements we derive a very accurate column density for Cyg X-1
of  6.21$\pm0.22 \times 10^{21}$ cm$^{-2}$. The results indicate that there are abundance
variations for Ne, O and Fe of the order of 10 to 20$\%$ relative to solar values; the deficiencies in
Fe and O are roughly consistent an abundance distribution of the interstellar medium. The Fe L edges
show narrow absorption line structure and for the first time were accurately modeled using most
recent laboratory data. These narrow line patterns could have given rise to residuals in the soft band
of previous observations of X-ray binaries. The spectral continuum is consistent with the models
applied to recent \xte spectra. It also shows a complex pattern of emission and absorption
lines. The observations do not allow to uniquely conclude on the origin of these discrete lines.

\centerline{\epsfxsize=8.5cm\epsfbox{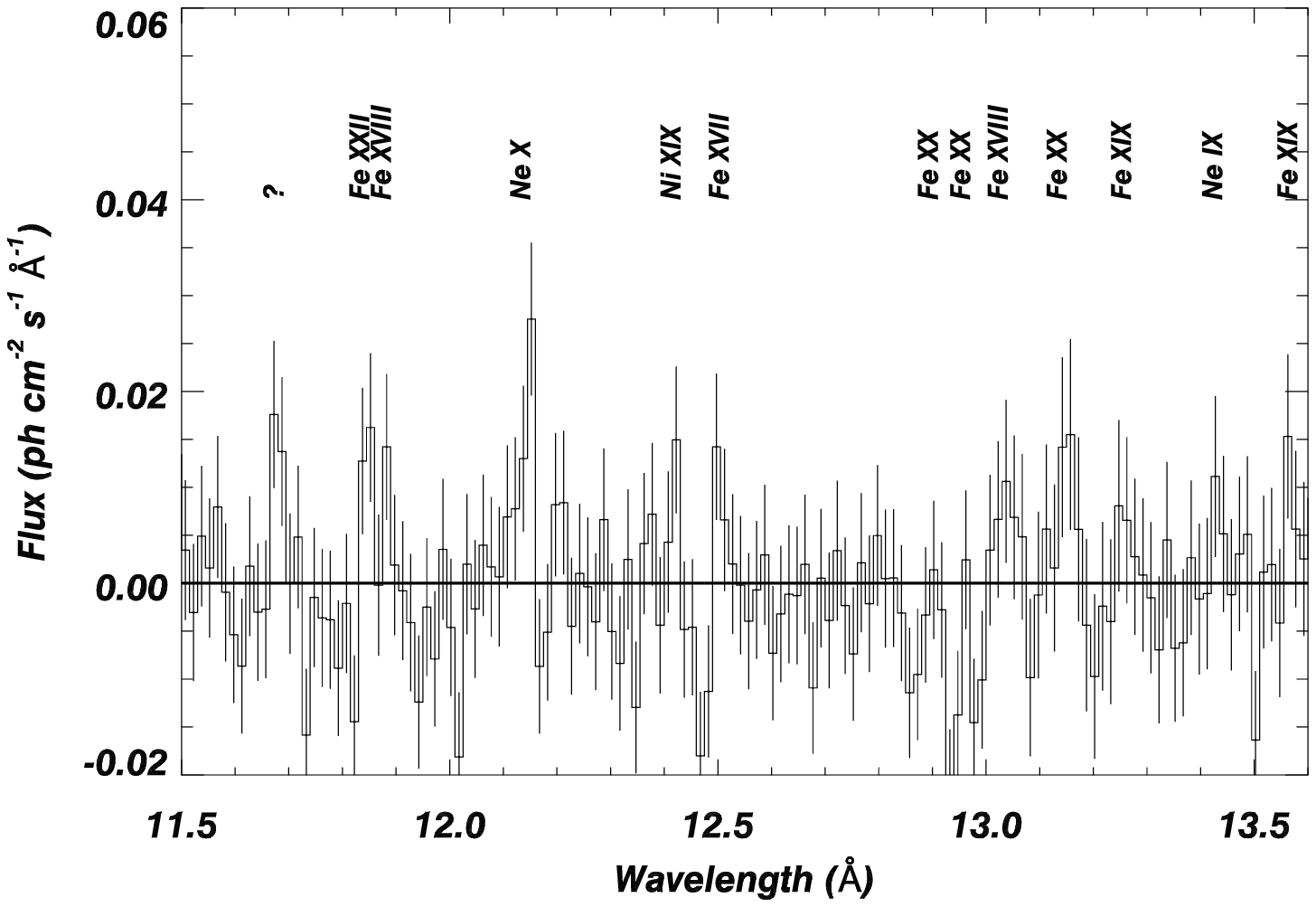}}
\vspace{-50mm}
\figcaption{A portion of the HEG spectrum between 11.5 and 13.6 ~\AA~. Candidate emission and absorption
line identifications are dominated by Fe. All marked features are consistent with the MEG response
function.
\label{nek}}
\vspace{2mm}

One of the most immediate benefits of the high spectral resolution
in the soft band is that we can measure the amount of photo electric absorption
almost independent of the intrinsic continuum.
A column density of 6.2$ \times 10^{21}$ cm$^{-2}$
was already suggested by Baluci\'nska and Hasinger (1991) from \exo studies of the soft excess in Cyg X-1.
The column density is expected to vary with
the phase of the orbital motion because the amount of stellar wind that
the line-of-sight passes through varies (e.g., Wen, Cui, \& Bradt 2000).
This might explain the lower values obtained by previous studies (e.g.,
Cui et al. 1998, Ebisawa et al. 1986, Remillard and Canizares 1984).
Using the most updated ephemeris
of Cyg X-1 (Brocksopp et al 1999b), we found that our observation corresponds
to an orbital phase 0.92, which is very close to the superior conjunction
(defined when the phase=0). In the discussion on the soft state by Wen et al. (2000)
it was argued that absorption at this orbital phase should be much higher than
the observed 6.2$ \times 10^{21}$ cm$^{-2}$, perhaps indicating a reduction
in absorption caused by high level of ionization of the wind in the flaring
state. On the other hand we note, that the derived column density is also quite
close to the expected value from the galactic HI distribution
(Dickey and Lockman 1990).

Perhaps the most detailed study of neutral absorption and abundance variations in Cyg X-1 spectra was
performed by Ebisawa et al. (1986). In several \asca observations, column densities between
5.3 and 6.0 $ \times 10^{21}$ cm$^{-2}$ were observed as well as an under abundance of oxygen of $\sim 20\%$ and
an over abundance of neon of $\sim 40\%$ compared to solar (as adopted by
Morrison and McCammon 1983). The {\em Chandra} observations confirm that there are abundance variations,
but at only a $\sim 10\%$ level for neon and oxygen.
We note that these abundance variations are probably unrelated to abundance variations observed
in intrinsically absorbed reflection components as suggested by Done and Zycki (1999),
to which our Chandra spectra are not sensitive.
In fact, measurements of the Fe K edge (Kitamoto et al. 1984, Marshall et al. 1993) can be
used to argue that here Fe is overabundant by a factor od two.

\centerline{\epsfxsize=8.5cm\epsfbox{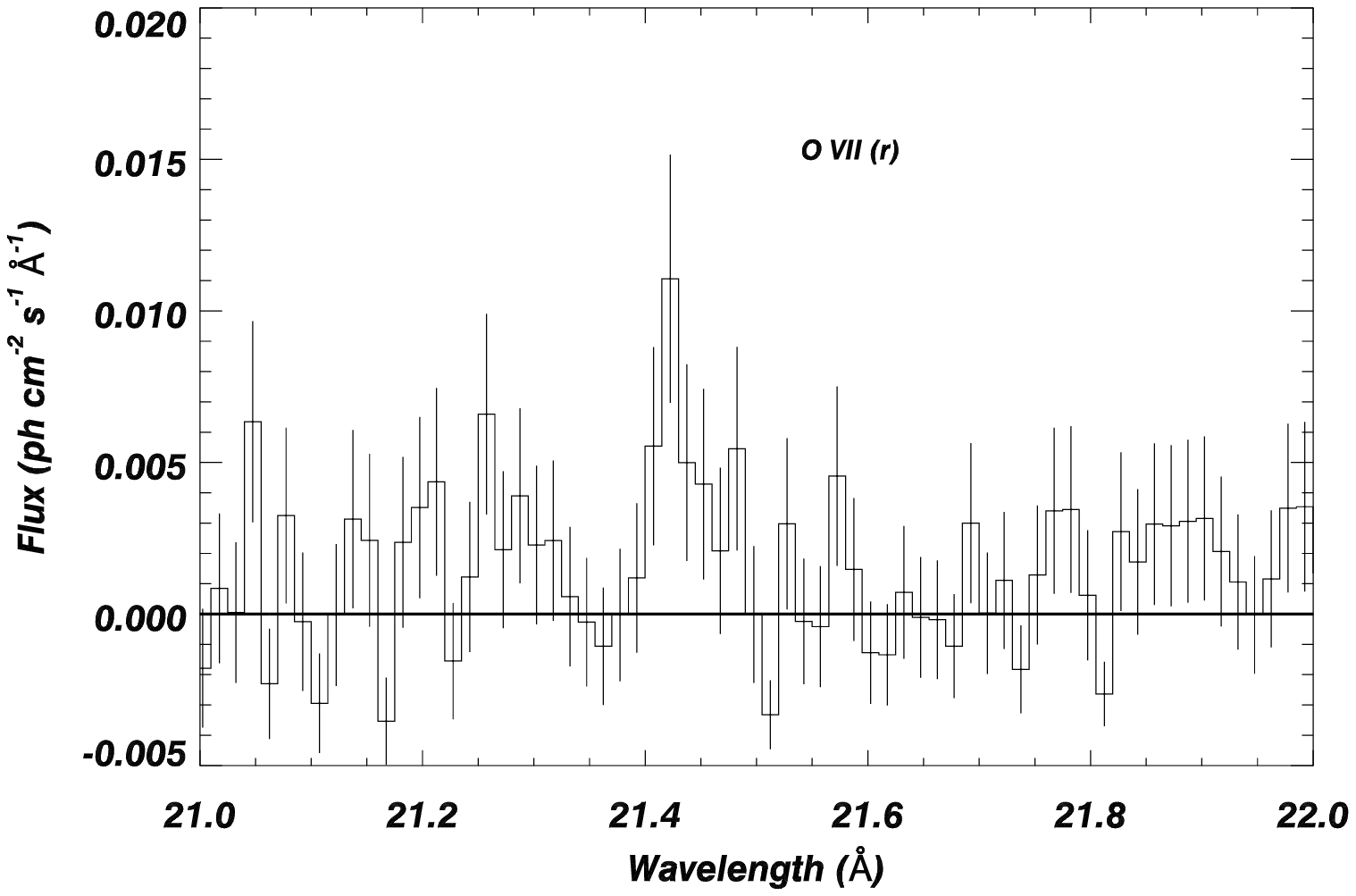}}
\vspace{-50mm}
\figcaption{A portion of the MEG spectrum showing a possible detection of a blue-shifted O VII line.
\label{nek}}
\vspace{2mm}

Barr and van der Woerd (1990) found
an emission line component at 730 eV with an eqivalent width of 370 eV, which was confirmed by Ebisawa et al. 1986,
but only as an alternative solution to the abundance variations. 
Our Chandra observations do not
show such a strong line feature below 1 keV and the solution of this puzzle clearly lies in the observed
structure of the photoelectric Fe L- and O K edges. The narrow structures we observe in the Fe L- and O K edges 
could give rise to residuals in the soft band in observations using low resolution detectors, which ultimately would be
interpreted as line features.
The detailed modeling using 
recent laboratory measurements (Kortright and Kim 2000) is quite successful in 
explaining the overall appearance of the Fe L edges due
to mostly metallic Fe with maybe some contribution from oxides. There is still an outstanding
issue in the model, an 8$\%$ effect, in that the shape of the Fe L3 edge is not entirely
explained by the new cross sections. However, we determine a specific column density due to
Fe of 1.55$\pm 0.12 \times 10^{17}$  cm$^{-2}$ indicating an under abundance of 25$\%$ with respect to solar.     
Such an iron depletion may also indicate that some material is bound in large grains (Predehl and Schmitt 1995). 
On the other hand, in Wilms, Allen, and McCray (2000) a large number of elements representing
the interstellar medium are listed slightly underabundant compared to a solar distribution.
The O K edge also shows extensive substructure and most of the details are consistent with
the edge in X0614+091 (Paerels et al. 2001). 

The determination of the total amount of neutral absorption
is essential for the modeling of the intrinsic continuum. Many broadband models have been
applied to Cyg X-1 X-ray spectra in the 1 and 10 keV X-ray band (Cui et al. 1998, 
Cui et al. 1997, Ebisawa et al. 1996) and most of them involved a soft blackbody or disk 
blackbody plus a power law. The studies based on low-resolution data
always suffer from the inability to disentangle various spectral
components, so the results tend to be model dependent. Now, the superior
spectral resolution of the HETGS allows us to measure the absorption edges accurately. From
the edges alone, we derived a reliable column density, independent of 
continuum modeling. With this reduction of free parameters, we were able to model
the continuum much more reliably. The results indicate that the source was
in a state that is similar to the transitional period between the hard and
soft states. This is consistent with the fact that the source was  observed during a period
of strong flaring.
 
The detection of modulations caused by the orbital period in the light curves and
hardness ratios of Cyg X-1 (e.g., Wen et al. 1999) can be interpreted as being caused
by a varying absorption optical depth through the companion wind with the binary motion. 
Gies and Bolton (1986) estimated that the terminal velocity of the wind is $\sim 1500$ km s$^{-1}$ based
on the radiatively driven wind model of Castor, Abbott, and Klein (1975).
Recent \chandra observations
of stellar winds from massive O-stars (Schulz et al. 2000, Kahn et al. 2001, Waldron and
Cassinelli 2001) find strong X-ray line emission with luminosities 
quite similar of the faint line features we observe in the Cyg X-1 spectrum and 
we find it very likely that some of these are signatures from the companion wind.
The  P Cygni type appearance of some of the lines bolster the case for a wind.
P Cygni X-ray lines have been observed in the spectra of Cir X-1
(Brandt and Schulz 2000) as well as, - tentatively -  in Cyg X-3 (Liedahl et al. 2000). In Cir X-1
the profiles are thought to originate in a high velocity disk wind, while in Cyg X-3
the extremely strong wind of the Wolf-Rayet companion star may generate such lines. 
Previously, Pravdo et al. (1980) observed wavelength shifts in IUE observations of HD226868, the companion
of Cyg X-1. Extreme blue shifts of Si IV and C IV lines up to
-1570 km s$^{-1}$ and -2270 km s$^{-1}$ were attributed to the stellar wind of
the companion. However, recent \chandra observations of massive O stars have
detected much smaller blueshifts
(Schulz et al. 2000, Waldron and Cassinelli 2001, Kahn et al. 2001, Cassinelli at al. 2001). 
It is therefore difficult to associate the derived -2300 km s$^{-1}$ shift of the O VII line with the stellar wind.
For Cyg X-1 it seems more likely that we observe
line emission from various different emission regions, which produce a complex pattern of emission and absorption in the X-ray spectrum. 
Future high resolution observations 
covering different orbital phases and with longer exposures may be used to identify the 
features that vary with the orbital motion and thus shed light on their
physical origin.

\acknowledgments

The authors want to thank the MIT HETG team 
and the Chandra X-ray Center for its support.
This research is funded by contracts SV-61010 and NAS8-39073.
WHGL also gratefully acknowledges support from NASA.
W.C. has also been supported by NASA through the LTSA grant NAG5-9998.

\end{document}